\documentclass[12pt, journal, draftcls, onecolumn]{IEEEtran}
\usepackage{algorithm}
\usepackage{algorithmic}
\usepackage{amsfonts}
\usepackage{amssymb}
\usepackage{cite}
\usepackage{color}
\usepackage{multirow}
\usepackage{subfigure}
\usepackage{array}
\usepackage{graphicx,times,amsmath}

\begin{document}

\title{KPI/KQI-Driven Coordinated Multi-Point in 5G: Measurements, Field Trials, and Technical Solutions}

\author{
	Guochao~Song,
	Wei~Wang,~\IEEEmembership{Member,~IEEE},
	Da~Chen,
	Tao~Jiang,~\IEEEmembership{Senior~Member,~IEEE}
	\thanks{G. Song is with the Wuhan National Laboratory for Optoelectronics, Huazhong University of Science and Technology. E-mail: \ sgc@hust.edu.cn.}
	\thanks{W. Wang, D. Chen and T. Jiang are with the School of Electronic Information and Communications, Huazhong University of Science and Technology. E-mail: \{weiwangw,chenda,taojiang\}@hust.edu.cn.}}

\maketitle

\begin{abstract}

The fifth generation (5G) \textbf{systems are} expected to be able to support massive number of wireless devices and intense demands for high data rates while maintaining low latency. Coordinated multipoint (CoMP) is advocated by recent advances and is envisioned to continue its adoption in 5G to meet these requirements by alleviating inter-cell interference and improving spectral efficiency. The higher requirements in 5G have raised the stakes on developing a new CoMP architecture. To understand the merits and limitations of CoMP in 5G, this article systematically investigates evaluation criteria including key performance indicators (KPIs) and key quality indicators (KQIs) in 5G, conducts empirical measurements and field tests, and then proposes a KPI/KQI-driven CoMP architecture that fulfills KPI requirements and provides KQI guarantee for each user. 

\end{abstract} 


\section{Introduction}

Recent years have witnessed a boom in mobile devices that are driving the current mobile communication systems toward their limits. To meet the requirements of the explosive growth wireless data traffic, the upcoming fifth generation systems are envisioned to employ large-scale multi-input multi-output (MIMO) technology to improve spectral efficiency\cite{5GCoMP,Rose}. By exploiting the space dimension, MIMO can support more connected devices with higher peak rates.

However, the real-world performance of the non-cooperative large-scale MIMO technology is limited by the issue that cell-edge users suffer inter-cell interference from adjacent cells\cite{5GCoMP,inteferenceCoMP}. To overcome this challenge, coordinated multipoint (CoMP), a distributed MIMO technology, is proposed by the 3rd generation partnership project (3GPP) in long-term evolution-advanced (LTE-A) to improve the performance of cell-edge users by scheduling multiple coordinated base stations (BSs)\cite{Realse9,Realse11}, and is expected to continue its adoption in the future 5G systems\cite{5GCoMP}.
 
Deploying a practical CoMP system for the cellular network is non-trivial. Backhaul network, imperfect channel state information (CSI), and synchronization substantially compromise the performance in CoMP, which degrades the key performance indicators (KPIs) and key quality indicators (KQIs) in 5G. The KPIs are certain quantitative parameters to evaluate network quality, and typical KPIs include peak rates and latency. The KQIs define the quality perceived subjectively by the end user, namely, quality of experience (QoE)\cite{KQIdefine}. They include the total KPIs, but are affected by confounding factors, such as user's preference and hardware diversity. Most studies have focused on the physical layer (PHY) and medium access control layer (MAC) to satisfy the KPIs requirements, while how to design CoMP schemes to meet KQIs in the upper layers has not yet been systematically investigated. Since future traffic demands are occupied significantly by visual-experience-oriented services\cite{qoesurvey}, such as panoramic videos, virtual reality (VR), and augmented reality (AR), KQI plays a crucial role in evaluating 5G technologies. Consequently, it is essential to take into account both KPIs and KQIs when designing CoMP schemes.

The goal of this article is to first investigate the key factors in CoMP affecting the KPI and KQI performance, and call attention to a clean-slate redesign of CoMP with KPI/KQI provisioning. Specifically, we start at a deep dive at the principles of downlink CoMP and shed light on the relationship among CoMP design factors, KQIs and KPIs in CoMP. Then, we implement a prototype testbed and analyze measurement results for KPIs and KQIs. Based on field test results, we present a feasible solution for how to deploy the practical CoMP systems, and then propose a KPI/KQI-driven CoMP architecture with KPI/KQI provisioning.

\section{KPI/KQI for CoMP in 5G}
In this section, we first describe the basic principle of downlink coordinated multipoint. Then, we investigate how backhaul, CSI feedback, as well as clock synchronization, influence the KPIs and KQIs in CoMP. Finally, we study the relationship between KQIs and KPIs and show the KQI is the better metric to evaluate user's experience.

\begin{figure}[H]
	\centering
	\subfigure[]{	
		\label{CoMP_principle.a}	
		\includegraphics[width=3in]{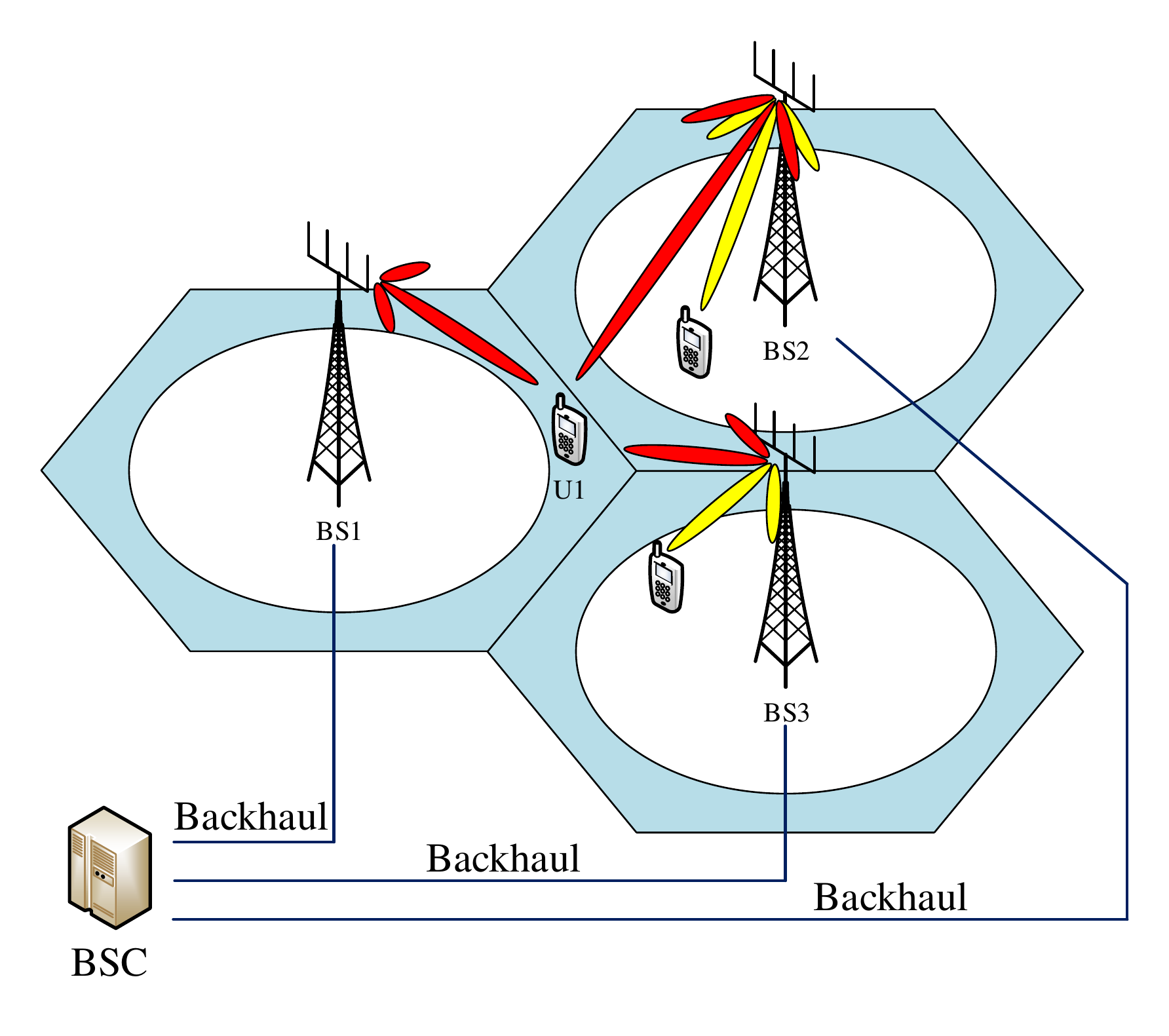}}
	\subfigure[]{	
		\label{CoMP_principle.b}	
		\includegraphics[width=3in]{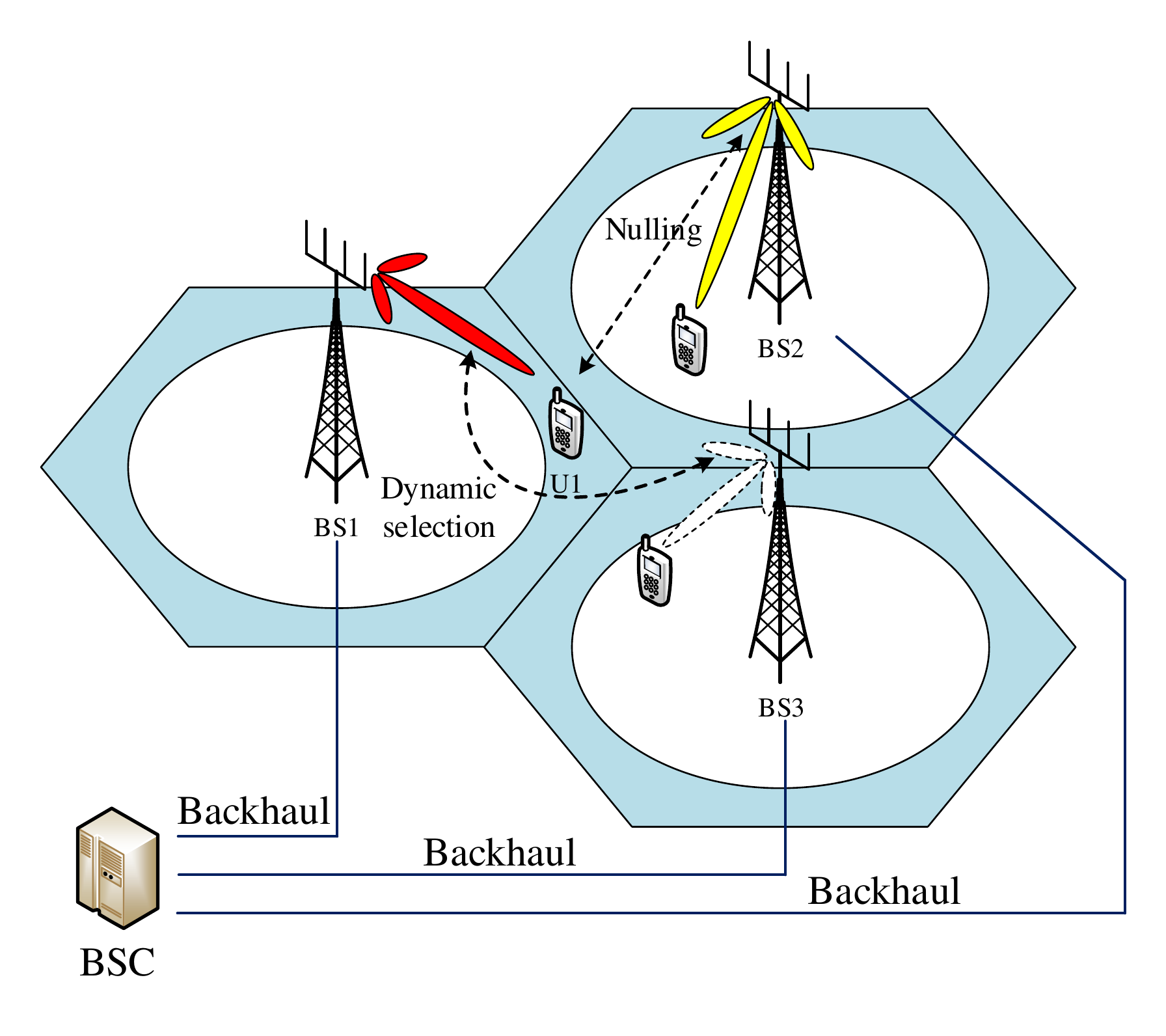}}
	\caption{\textit{The basic principle of the downlink CoMP: a) JT; b) CS/CB.}}
	\label{Fig.CoMP_principle}
\end{figure}

\subsection{Downlink Coordinated Multi-Point Transmission}

The upcoming 5G will have to support a multitude of new applications with a wide variety of requirements, including higher peak rates, reduced latency, increased number of devices, and so on\cite{5GCoMP}. To meet these requirements with limited spectral resources, improving spectral efficiency is of great significance. However, inter-cell interference is still a huge barrier for cellular networks to improve spectral efficiency. Recent research has demonstrated that CoMP has great potential to manage inter-cell interference and improve spectral efficiency\cite{inteferenceCoMP,downlinkCoMP}. As a result, CoMP is envisioned to continue its adoption in 5G\cite{5GCoMP}. In the framework of LTE, downlink CoMP transmission can be divided into a pair of categories, namely, joint transmission (JT) and coordinated scheduling/beamforming (CS/CB)\cite{downlinkCoMP}. In the following, we will describe the basic principle of the JT and CS/CB\ in detail.

As shown in Fig.~\ref{CoMP_principle.a}, by joint signal processing among coordinated BSs, JT can exactly pinpoint each user, and ensure that the signal from all coordinated BSs can be added up at user's position, while creating no interference to other users\cite{downlinkCoMP}. The joint signal processing is based on MIMO  precoding technology to support multiple concurrent data streams as much as possible\cite{filedCoMP,downlinkCoMP}. Additionally, as a distributed technology, JT can achieve extra degrees-of-freedoms (DoFs) with rich propagation paths, thus serving more users in the same time-frequency resource. Despite the fact that JT can achieve high capacity gain by spatial reuse, deploying a practical JT system is non-trivial. The main challenges include three aspects. First, all joint signal processing is implemented in a base-station controller (BSC), and thus it is necessary to exchange channel state information (CSI) and data for the target user to the BSC. Due to the significant feedback overhead, the performance of JT is directly dictated by the backhaul network among coordinated BSs\cite{backhaulCoMP}. Second, the joint precoding matrix is derived by CSI, and thus the performance of JT heavily relies on the accuracy of CSI\cite{inteferenceCoMP}.
Finally, each BS is driven by independent local clocks, resulting in carrier frequency offset between BSs. With the frequency offset, signal phases from different BSs rotate at different speeds, thereby preventing joint signal processing.

CS/CB departs from JT in that it adopts traffic management mechanisms in which the BS nullifies its interference to the users of adjacent cell. The basic principle of CS/CB is illustrated in Fig.~\ref{CoMP_principle.b}. When BS1 is serving U1, BS2 needs to nullify its interference to U1 by using a DoF of beamforming. Moreover, to avoid interfering U1, BS3 is scheduled to serve U3 in a different time slot. Therefore, in CS/CB, the BSs need to sacrifice they resources in space domain or time domain to nullify they interference to the users of adjacent cell. The performance of CS/CB is bound up with the accuracy of CSI and the performance of backhaul network. The CSI decides the directions of cell-edge users and the beamforming, and inaccurate CSI drastically degrades the performance of CS/CB. The backhaul network needs to exchange CSI and indirectly affects the performance of CS/CB.

\subsection{KPI Metrics in CoMP Systems}
\begin{figure}[t]
	\centering
	\includegraphics[width=5.5in]{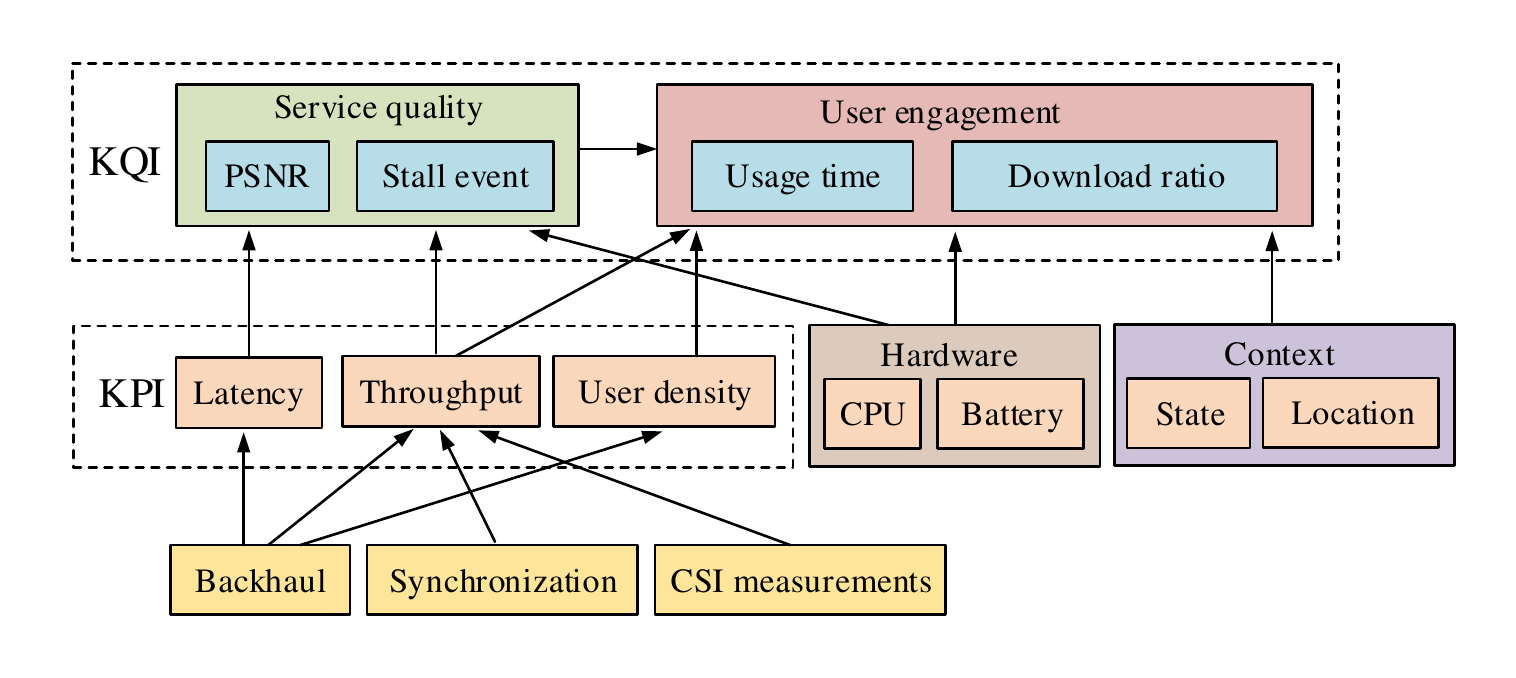}
	\caption{\textit{The complex interdependencies between KQI and KPI.}}
	\label{fig:CoMP_metric}
\end{figure}

KPIs are some key quantitative parameters that define user's quality of service\cite{KQIdefine}. Typical KPIs in the 5G contain peak rates, end-to-end latency, the rate of cell-edge users, as well as connection density. To achieve these very challenging KPIs, CoMP is an irreplaceable technology in 5G. Indeed, for the practical CoMP systems, limited backhaul and imperfect CSI and synchronization errors prevent CoMP to attain full theoretical gains, therefore limiting the KPIs in CoMP. Next, we investigate how the key factors influence the KPIs, respectively.

\subsubsection{Backhaul}
The backhaul needs to exchange data to coordinated BSs, including the downlink CSI and raw in-phase/quadrature (I/Q) signals for each user. When the throughput of backhaul is insufficient, there are two approaches to reduce the overhead of exchanging data: quantizing the CSI and I/Q signals with fewer bits, and reducing the number of connected users. As a result, the throughput will degrade with higher quantizing noise, and the user density will also decrease with less connected users. Additional, the delay of exchanging data is significant, which introduces extra latency to each user. 5G will deploy more antennas in BSs to improve the spectral efficiency, which further enhances the demand for backhaul. When the throughput of backhaul cannot meet the demand of JT, CS/CB without exchanging I/Q signals will be a suboptimal solution. To sum up, the limited backhaul drastically degrades the KPIs in CoMP, including throughput, end-to-end latency, and user density.

\subsubsection{Channel State Information}
Imperfect CSI seriously reduces the throughput of CoMP\cite{downlinkCoMP}. Indeed, it is non-trivial to obtain accurate and real-time CSI. In LTE Release8, the reported CSI is a set of transmission parameters recommended by user corresponding to a transmission hypothesis\cite{inteferenceCoMP}. The rank indicator (RI) is the rank of channel matrix or number of DoF. The precoding matrix index (PMI) denotes an index of a predefined precoding matrix. The channel quality indicator (CQI) is a reference for adaptive rate selection. Imperfect CSI makes RI and PMI, as well as CQI, be inaccurate. As a result, the coordinated beamforming and joint signal transmission cannot eliminate inter-cell interference. In JT, users connect with all coordinated BSs, so propagation paths from BSs to user are richer. Thus, JT can simultaneously support more users by spatial multiplexing. However, measuring CSI occupies massive feedback overhead in the uplink. To reduce feedback overhead, CSI has to be compressed in multiple dimension such as feedback frequency and the number of
quantization bits, which degrades the throughput of CoMP. 

\subsubsection{Clock Synchronization}
Clock synchronization at the coordinated BSs side plays an important role in CoMP with the distributed MIMO architecture, since the coordinated BSs cannot share the same timing and frequency resources. Clock synchronization includes time synchronization and frequency synchronization. Time synchronization error will cause inter-symbol interference (ISI), and frequency synchronization error will cause inter-carrier interference (ICI). The coordinated BSs need to transmit the data frames on the assigned time slots in order to avoid time synchronization error, which is an enormous challenge for the current infrastructures.  Fortunately, in LTE-A, the primary synchronization signal (PPS) and the secondary synchronization signal (SSS) are available for reference timing, and the evaluation in 3GPP has shown the timing error is less than 3~$\mu s$. However, frequency offset between coordinated BSs is still an enormous barrier. With the frequency offset, the signal phases from different BSs rotate at different speeds, thereby causing ICI. The ICI will lead to significant throughput degradation in JT.

\subsection{KQI Metrics in CoMP Systems}

Backhaul network, clock synchronization, and CSI measurements directly affect KPIs and also indirectly influence  KQIs. Compared with KPIs, KQIs are some qualitative metrics that quantify QoE\cite{KQIdefine} and can reflect user's perceived experience. As shown in Fig.~\ref{fig:CoMP_metric}, KQIs depend not only on KPIs but also on other factors, such as hardware diversity and context differences. Indeed, CoMP is a conventional physical layer (PHY) design and aims to realize these demands of KPIs in 5G, while it cannot provide KQIs guarantee for each user's perceived experience. To fully understand CoMP from the KQIs' perspective, we study the relationship between the KQIs and KPIs in CoMP on as follows.

KQIs include a pair of metrics, that is, service quality and user engagement. Note that, for different applications, the KQI metrics are distinct. To be succinct, we list several essential performance just for the visual applications, such as AR, VR, and panorama videos. Both peak signal-to-noise ratio (PSNR) and stall event are the most common metrics to quantify the performance of video reconstruction. PSNR is used to quantify the performance of each video frame, and it depends on the throughput of CoMP. Stall event is also a crucial indicator, which represents whether the video can play smoothly. When there exist high latency and low throughput in CoMP, the video playing will stall. PSNR and stall event are also affected by hardware, such as CPU processing ability, battery status, and memory cache. User engagement is a more actionable metric to evaluate user's satisfaction with the video service\cite{qoesurvey}. However, user engagement depends not only on service quality but also on the user service context and hardware capabilities. Recall that backhaul network, clock synchronization, and CSI measurements directly affect KPIs. Thus, these factors also indirectly influence KQIs. 



\section{Measurements and Field Trials}

In this section, we first test the key performance of CoMP in field trials and show the relationship between the KPIs of CoMP and various factors, including backhaul, CSI feedback interval, as well as synchronization. We then evaluate the KQI of CoMP and demonstrate that KPI-driven CoMP cannot provide guarantee for each user's experience.

\subsection{Field Trials for CoMP}

\begin{table}[t]
	\centering
    \caption{\textit{The parameters of the field trial.}}
	\begin{tabular}{|l|l|}\hline
		Parameter&Value\\\hline
		The number of BSs&3\\
		Antenna configuration(user, BS)&(1, 4)\\
		Center frequency&3.5 GHz\\
		Sampling rate&30.725 MHz\\
		Active subcarriers&1200\\
		Transmission Power&20 dBm\\
		Subcarrier interval&312.5 kHz\\
		Modulation&2PSK/4QAM/16QAM\\
		Channel Coding&LDPC with 1/2 code rate\\
		Duplex mode&FDD\\\hline
	\end{tabular}
	\label{table1}
\end{table}

A prototype testbed was built and a field test was performed on a campus. The testbed includes three four-antenna BSs and a large number of single-antenna users. To obtain enough data, we repeat test for users at various locations in indoor environments. The parameter settings of the physical layer for the prototype testbed are listed in TABLE~\ref{table1}. The clock sampling rate of the baseband is set as 30.725 MHz, and the carrier frequency is 3.5 GHz. Orthogonal frequency division multiplexing (OFDM) is used for the PHY modulation scheme, and the number of active subcarriers is 1200. The testbed employs 2PSK/4QAM/16QAM and low-density parity-check code (LDPC) with 1/2 code rate as modulation and coding scheme, respectively. The testbed uses frequency division duplex (FDD), and the CSI of the downlink channel is fed back to BS by the uplink. In addition, PXIe bus is designed for the backhaul network, and its maximum throughput is 240 Gb/s. The reference clocks of BSs are connected with the same clock distribution module resource by SMA cable, and the clock distribution module can supply 8-channel clock with 10 MHz. In the following, we evaluate how backhaul, CSI, and synchronization affect the KPIs in CoMP.

\begin{figure}[t]
	\centering
	\subfigure[]{	
		\label{Field_test.a}	
		\includegraphics[width=3.3in]{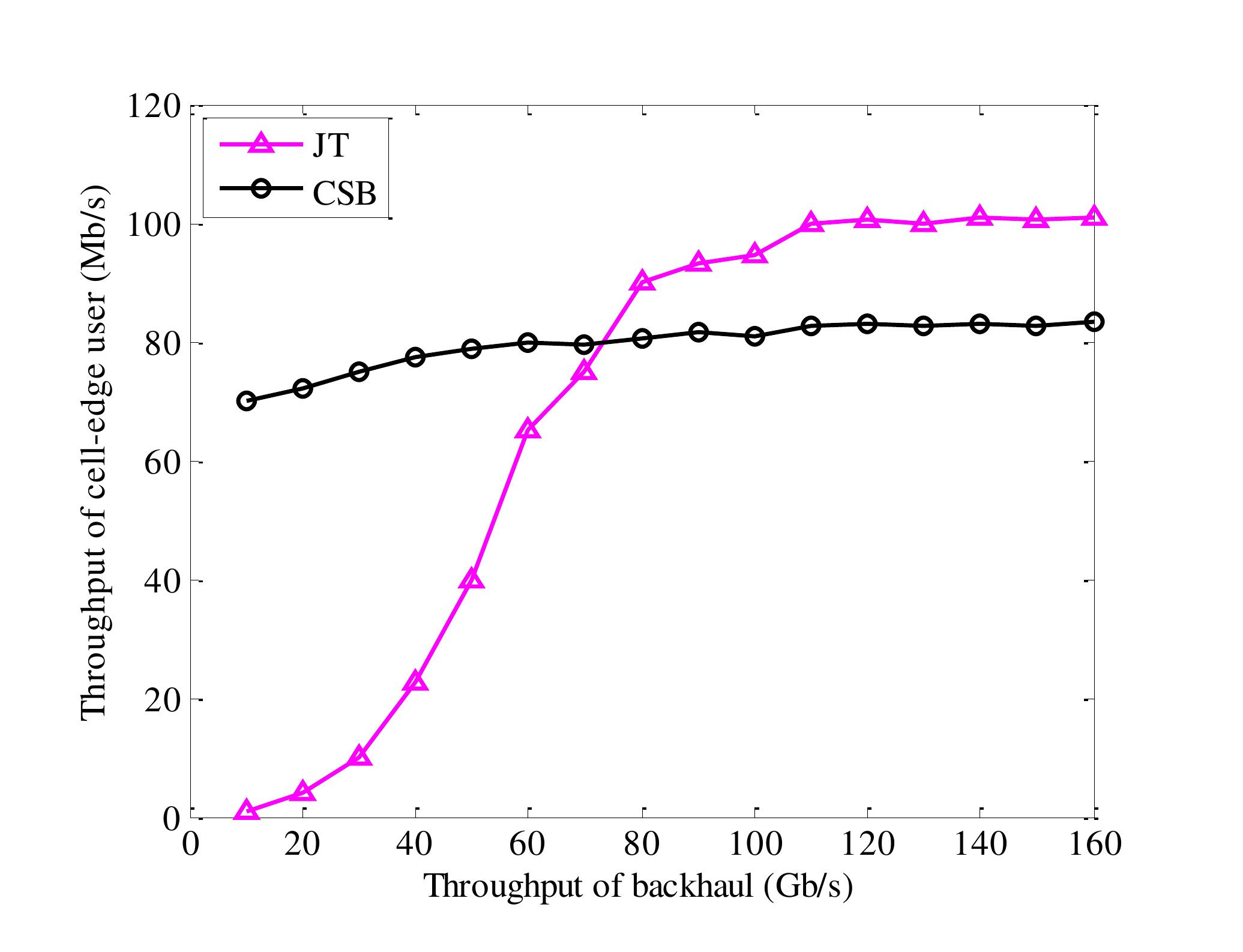}}
        \hspace{-.5in} 
	\subfigure[]{	
		\label{Field_test.b}	
		\includegraphics[width=3.3in]{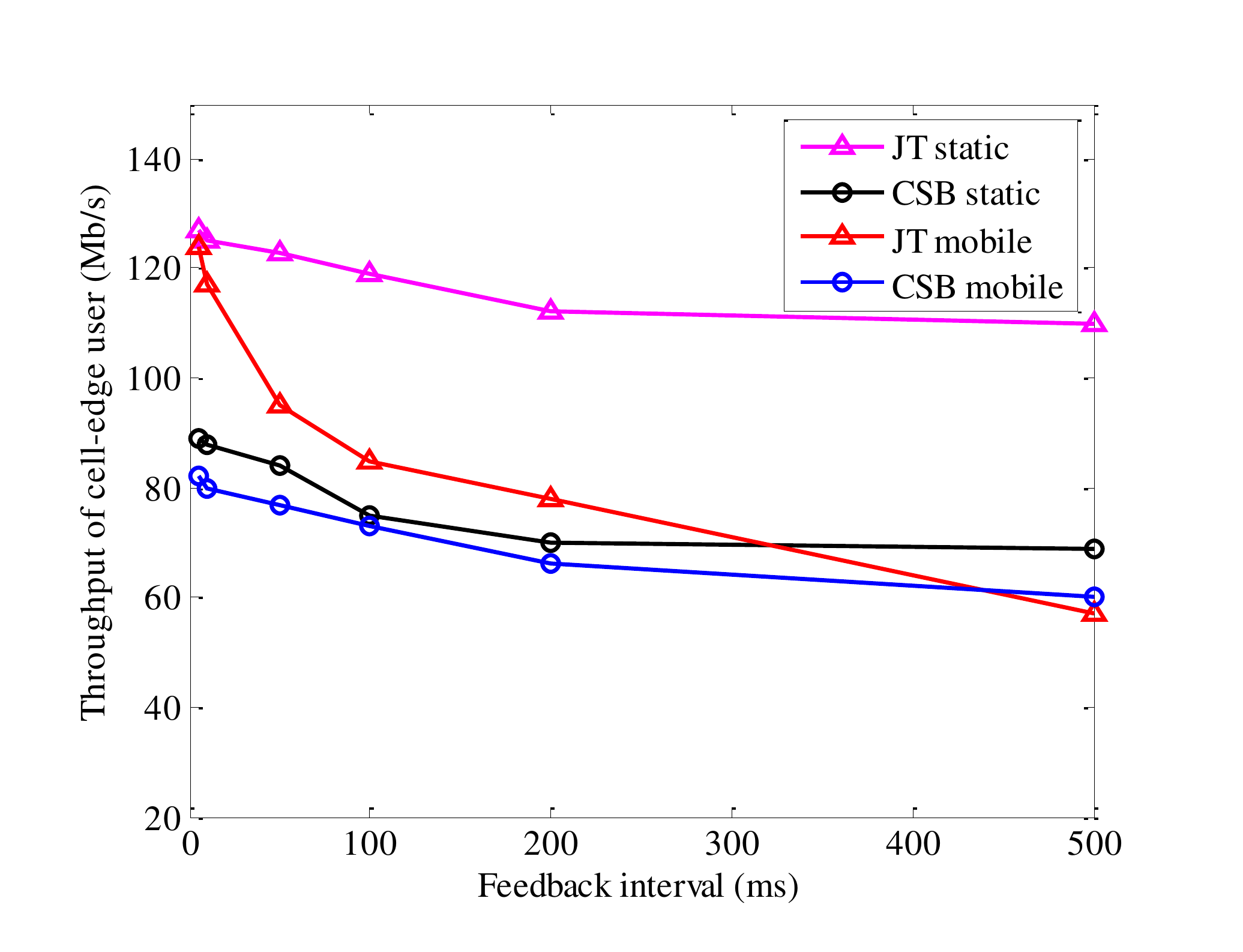}}
 
	\subfigure[]{	
		\label{Field_test.c}	
		\includegraphics[width=3.3in]{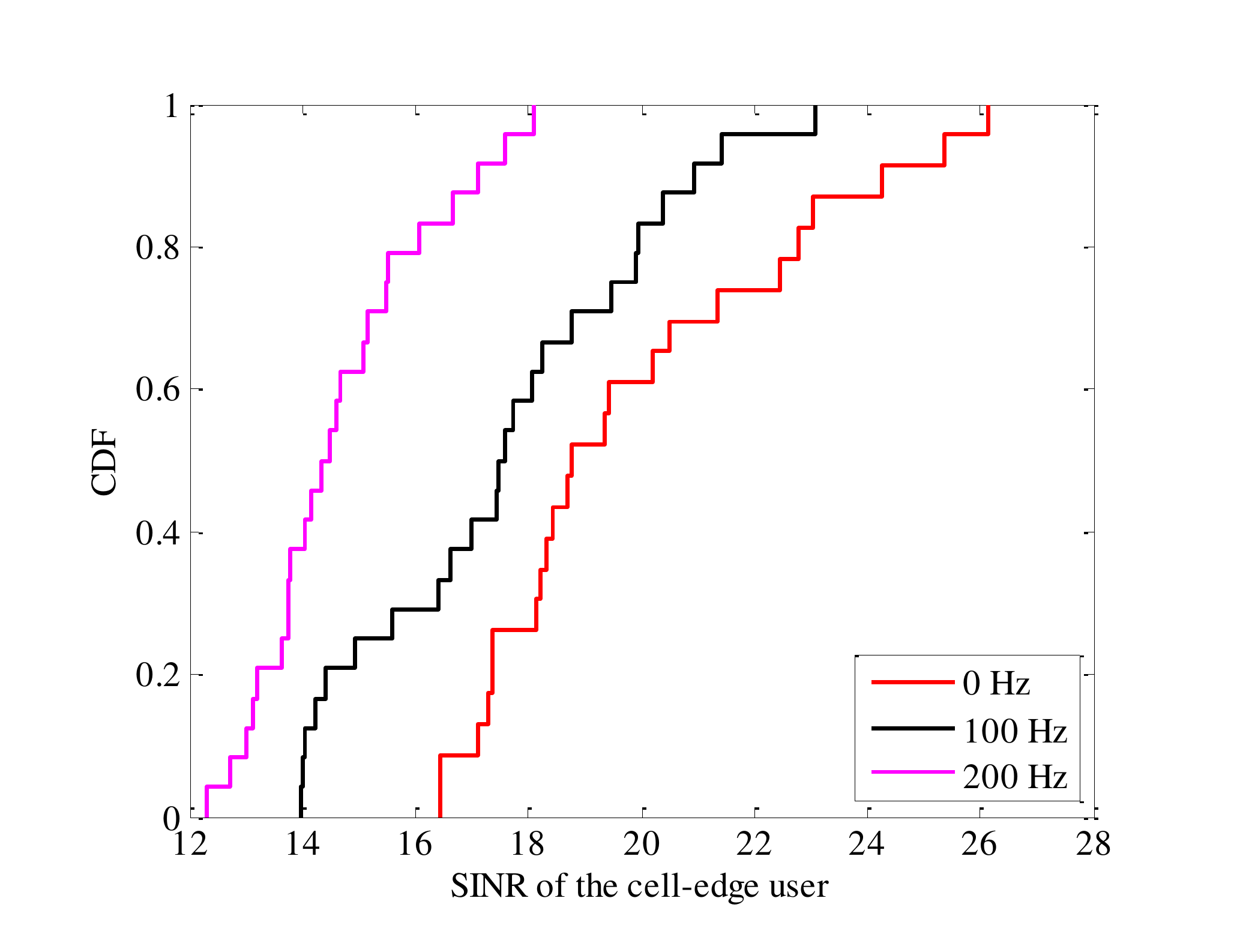}}
	    \hspace{-.5in} 
	\subfigure[]{	
	\label{Field_test.d}	
	\includegraphics[width=3.3in]{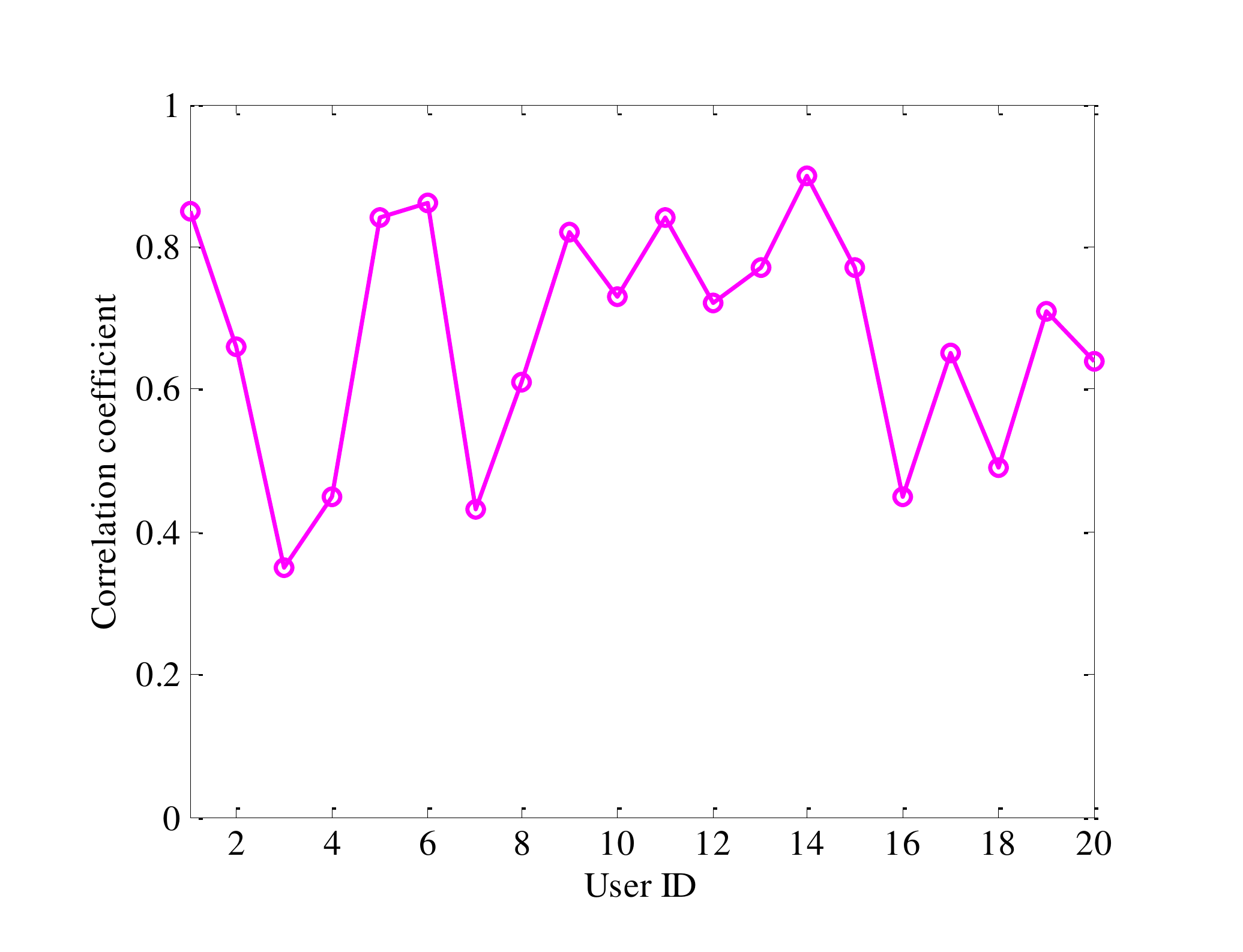}}	

	\caption{\textit{Field trials and measurements: a) The throughput of cell-edge user vs. throughput of backhaul; b) The throughput of cell-edge user vs. channel feedback interval; c) The SINR vs. frequency offset; d) The correlation between download ratio and downlink throughput for 20 random users.}}
	\label{Field_test}
\end{figure}

\subsubsection{Backhaul}

Recall that the performance of JT largely depends on the backhaul network. To explore how backhaul network influences JT and CS/CB, we test the throughput of the cell-edge user. In the test, we gradually increase the throughput of backhaul network by controlling the maximum throughput of PXIe. Note that the throughput of backhaul network decides the quantization accuracy of the raw I/Q signal and CSI. When the throughput of backhaul network is limited, the raw I/Q signal and CSI have to be quantized with fewer bits, and vice versa. Fig.~\ref{Field_test.a} shows the throughput of the cell-edge user over throughput of backhaul. It is obvious that the performance of JT dramatically degrades when the throughput of backhaul network is inadequate. The reason behind this phenomenon is that JT needs to exchange raw I/Q signal, resulting in significant overhead. When the throughput of backhaul reaches 80 Gb/s, the performance of JT is beginning to stabilize. Compared with JT, CS/CB only exchanges CSI, and thus it can maintain the better performance when the throughput of backhaul network is limited. 

\subsubsection{CSI Feedback}
Both JT and CS/CB require the BS to know the downlink channel before transmission. In the absence of CSI, the CoMP cannot perform coordinated actions across multiple BSs. In the following, by leveraging the prototype testbed, we evaluate the performance of CoMP with different measurement schemes. Fig.~\ref{Field_test.b} shows the throughput of the cell-edge user for various feedback intervals of CSI in mobile and static scenarios, respectively.  As we can see, the throughput drastically decreases with the feedback interval, especially in the mobile scenario. Indeed, reducing the feedback interval can effectively improve the throughput of the downlink, but it occupies most of the resources of the uplink. Thus, it is worth considering how frequently CSI should be refreshed to balance the tradeoff between the throughput of the downlink and the feedback overhead of the uplink.

\subsubsection{Clock Synchronization}
Clock synchronization deviations preclude joint signal processing for JT. Thanks to CP in OFDM, time synchronization error can be tolerated. However, frequency synchronization is still a barrier, and slight frequency offset between coordinated BSs introduces serious ICI. As shown in Fig.~\ref{Field_test.c}, when the frequency offset becomes high gradually, the signal-to-interference-plus-noise-ratio (SINR) in JT declines rapidly. However, it is unnecessary to exchange raw I/Q signal for CS/CB, so that CS/CB can immune the frequency offset. In LTE, the reference clock is supplied by global position system (GPS), and its clock accuracy is from 20 ppb to 75 ppb. With the 3.5 GHz center frequency of RF, the frequency offset is from 70 Hz to 262.5 Hz. From Fig.~\ref{Field_test.c}, we can see that the slight frequency offset cannot be ignored. As a result, over-the-air synchronization among coordinated BSs needs to be performed to compensate the frequency offset, which is a common mechanism in wireless communication systems to make sure that different devices can share the common frequency and timing for data transmission.

In this section, we evaluate the KQI for CoMP system using data-driven simulations. To model the KQI requirements, we collected a large-scale data set from tier-one cellular service provider in China, which contains millions of subscribers and covers thousands of cells, including the complete traces in the core network and user profile database. We consider a scenario where three adjacent BSs form a cooperative cluster, and the BS transmission power is 20 dBm. The Multi-User MIMO (MU-MIMO) technology is used to serve these random distributed users. The criterion of scheduling for MU-MIMO is based on water-filling theory, whose goal is to maximize the throughput of the wireless network by resource allocation. We employ Sony Demo as video stream data, which is
compressed in MPEG4 with different quality levels. We utilize the user engagement to evaluate the user's satisfaction with the video service. The download ratio is adapted to quantify the user engagement, and it can be denoted as
\begin{equation*}
\label{ratio}
R = \dfrac{D}{S} \text{,} 
\end{equation*}
where $ D $ denotes the downloaded bytes, and $ S $ represents the video file size in bytes. To verify the existence of diversity on the user engagement, we randomly select 20 users and compute the Pearson correlation coefficient between download ratio and downlink throughput for each user, and it can be represented as
\begin{equation*}
	\label{correlation}
	\rho_{R,C}=\dfrac{E[(R-\mu_R)(C-\mu_C)]}{\sigma_R\sigma_C}\text{,} 
\end{equation*}
where $ C $ represents the downlink throughput, $ \mu_R $ and $ \mu_C $ denote the means of the download ratio and the downlink throughput, $ \sigma_R $ and $ \sigma_C $ are the standard deviations of the download ratio and the downlink throughput, respectively.  As shown in Fig.~\ref{Field_test.d}, the correlation coefficients of different users are very distinct, which reflects the user engagement is quite diversified. Thus, we show that conventional CoMP cannot provide KQI guarantee for each user.



\section{KPI/KQI-Driven CoMP Solutions}
 
In this section, based on filed trials, we first envision the feasible solution for deploying practical CoMP systems in 5G. We then propose a KPI/KQI-driven CoMP framework that can provide KPI/KQI guarantee for the CoMP systems.

\subsection{How to deploy practical CoMP system in 5G}
Based on our field trials, we analyze these challenges for deploying practical CoMP systems in 5G. Additionally, we envision some feasible solutions to overcome these challenges to provide the essential guarantee for KPIs and KQIs in CoMP systems.

\subsubsection{Backhaul}
The deployment of backhaul is one of the huge challenges for practical CoMP systems in 5G. BSs become denser and more Bss' antennas are set up in 5G. Thus, the backhaul network has to maintain higher throughput, otherwise the KPIs will be dramatically degraded in CoMP. The optical technology, such as wavelength division multiplexing (WDM), is the current solution for backhaul network. However, adopting optical technology as backhaul is very cumbersome. On the one hand, WDM needs essential relays to expand transmission distance, which is very tedious. On the other hand, when a new BS joins the cellular network, the new BS must establish the connection with all adjacent BSs and the topology of backhaul has to be rearranged, resulting in a lot of expenditures of workforce and resources. MmWave technology is an appearing solution of wireless backhaul network, which is cost-effective and scalable\cite{mmwave,5Gbackhaul}. However, mmWave cannot meet the full demand of backhaul network. Indeed, multiple adjacent BSs form a cooperative cluster and the sum throughput of backhaul of the cluster is constant. Our key observation is that CS/CB is a suboptimal solution when the throughput of backhaul network is inadequate. Thus, we can dynamically formulate the throughput of backhaul to each BSs of the cluster to achieve maximum performance gains.  

\subsubsection{CSI Feedback}
Reducing the overhead of CSI measurements is still an open and challenging issue in both FDD systems and TDD systems. Implicit feedback is an effective mechanism to address this issue. LTE Rel-11 employs implicit feedback for TDD systems, where downlink and uplink channels are considered to be reciprocal. Thus, BS can obtain the downlink channel by estimating the uplink channel. Recently, the implicit feedback is also used for FDD systems \cite{Dina}. The basic principle is that underlying physical paths stay same for different frequencies so that the downlink channel can be inferred by the uplink channel. To further reduce the feedback overhead, we propose an adaptive compression scheme, where the feedback interval is related to the channel coherent time. By estimating the correlation coefficient between two instances of the channel, we can infer the channel coherent time and dynamically adjust the feedback interval to balance the tradeoff between the feedback overhead of the uplink and the throughput loss of the downlink.

\subsubsection{Clock Synchronization}
Synchronization is also a fundamental requirement for distributed BSs in JT-CoMP. As shown in Fig.~\ref{Field_test.c}, the synchronization errors among coordinated BSs degrade the performance of JT-CoMP. Since the BSs' clock supplied by GPS is inaccurate, over-the-air synchronization is worth considering to help BSs to further improve synchronization accuracy. To this end, we can build a master-slave protocol, an over-the-air synchronization scheme, in which one BS acts as the master BS, and other BSs act as slaves. The slaves hear a beacon signal from the master BS to estimate and compensate the synchronization errors relative to master BS. Through the protocol, all BSs can obtain the same reference timing and frequency.

\subsection{KPI/KQI-Driven Scheduling in CoMP}

\begin{figure}[t]
	\centering
	\includegraphics[width=6in]{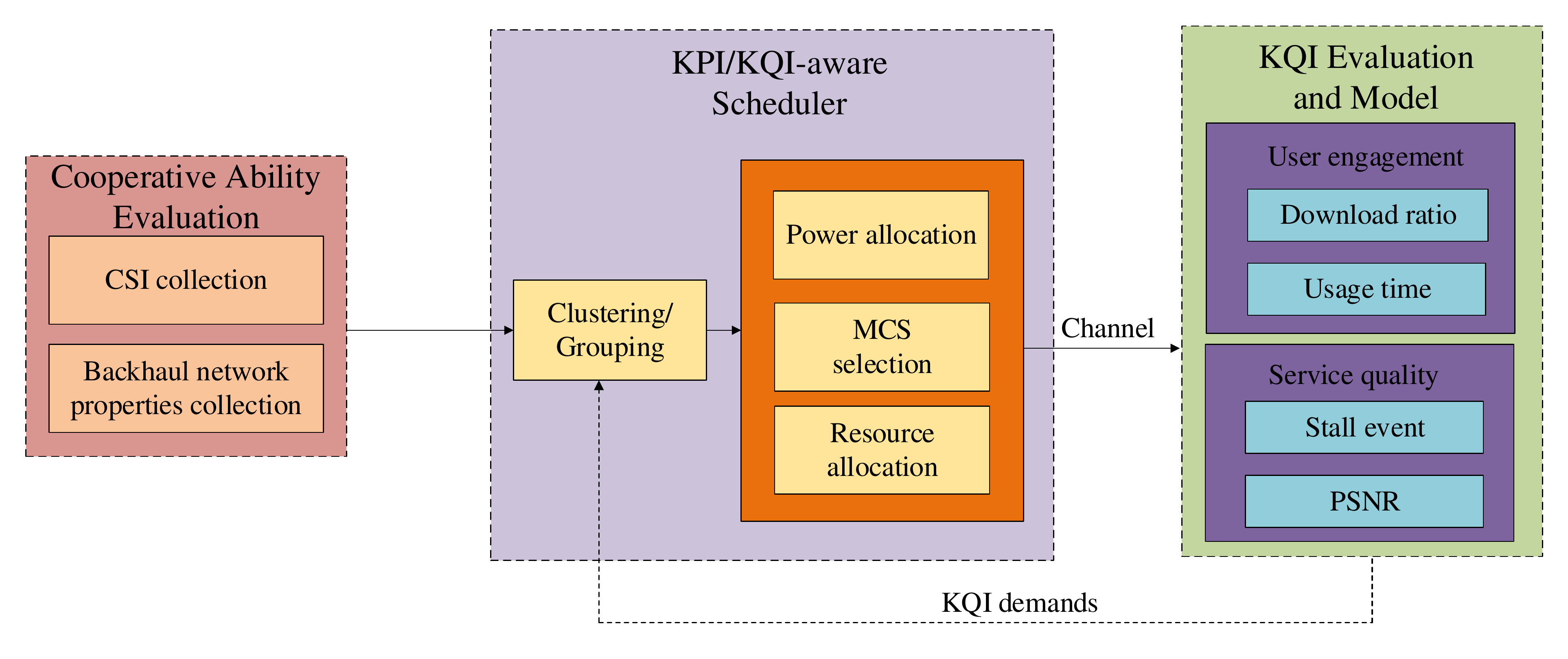}
	\caption{\textit{KPI/KQI-driven framework.}}
	\label{fig:CoMP_KQI_framework}
\end{figure}

We propose a cross-layer CoMP framework that provides KPI/KQI guarantee for the CoMP system. Fig.~\ref{fig:CoMP_KQI_framework} illustrates the proposed loop-closed KPI/KQI-driven framework that facilitates the CoMP system to improve user's perceived experience. 

\begin{itemize}
	
\item\textbf{KQI Evaluation and Model}. The aim of KQI evaluation is to evaluate KQI demand for each user based on the historical data and current data, including user engagement and service quality. By training these datasets, the KQI model can translate the KQI demand to system parameters for CoMP. These system parameters will be fed back to BSC by the uplink. 

\item\textbf{Cooperative Ability Evaluation}. The function of operative ability evaluation is to collect configuration information of BSs and CSI from all user in coordinated BSs. The configuration information is used to quantify the cooperative ability of BSs, which contains the throughput of backhaul, the number of antennas and transmit power. The CSI is used for the decision of cell-edge users and users grouping. 

\item\textbf{KPI/KQI-driven Scheduler}. The scheduler is a strategy maker in KPI/KQI-driven CoMP, whose goal is to provide the KQI guarantee as possible for each user, instead of maximizing the throughput of the wireless network in conventional CoMP. First, the scheduler performs BSs clustering based on the cooperative ability of each BS. When the cooperative ability is limited, the CS/CB is used as a suboptimal solution. The scheduler then groups users by the correlation between users' channel and the priority of users' data. Finally, the scheduler performs power allocation, modulation and coding scheme (MCS) selection and resource allocation. 

\end{itemize}
To provide KQI guarantee for each user in a long-term way, we formulate the resource allocation problem and clustering/grouping problem as a dynamic scheduling problem. The objective is to cluster BSs and group users in CoMP and optimize resource allocation to minimize time-averaged total KQI loss. However, due to users' mobility, the quality of wireless network significantly varies, thereby directly finding an optimal solution is difficult. Consequently, we have to divide this long-term scheduling problem into sub-problems for every time slot. Since BSs' DoFs are limited, BSs cannot serve all users in the same slot. Thus, we have to group users based on the priority of users' demands and the correlation of downlink channel between different users. Based on the Lyapunov drift-plus-penalty method in\cite{QueueingBook,Hu}, the objective is to maximize $\sum_u Q_u(t) \cdot b_u(t) - V \sum_u I_u(t)$. where $Q_u(t)$ is the data queue for user $u$ in slot $t$, $b_u(t)$ the video playing duration for user $u$ in slot $t$, $I_u(t)$ the KQI loss of user $u$ in slot $t$ and $V$ the control parameter of the drift plus penalty obtained from users' KQI preferences. Finally, we select the optimal resource allocation and best MCS to delivers services while providing KQI guarantee for each user.

\begin{table}[t]
	\centering
	\renewcommand{\multirowsetup}{\centering}
	
	\caption{\textit{The evaluation of correlation coefficient and number of stall event.}}
	\begin{tabular}{|p{2cm}<{\centering}|p{2.3cm}<{\centering}| p{2.3cm}<{\centering}|p{2.3cm}<{\centering}|p{2.3cm}<{\centering}|}\hline 
		
		\multirow{3}{2cm}{\textbf{Metrics}} & \multicolumn{2}{c|}{\textbf{Correlation coefficient}} &\multicolumn{2}{c|}{\textbf{Number of stall event}}\\
		\cline{2-5}
		{} &\textbf{KPI-driven CoMP} & \textbf{KPI/KQI-driven CoMP} & \textbf{KPI-driven CoMP} & \textbf{KPI/KQI-driven CoMP} \\\hline
		User 1 & {0.34} & {0.78} & {21} & {5}\\\hline
		User 2 & {0.55} & {0.89} & {15} & {3}\\\hline
		User 3 & {0.86} & {0.92} & {9} & {4}\\\hline
		User 4 & {0.52} & {0.79} & {31} & {11}\\\hline
		User 5 & {0.71} & {0.88} & {17} & {2}\\\hline
		
	\end{tabular}
	\label{table2}
\end{table}

We compare the KPI/KQI-driven framework with the KPI-driven framework where KQI demands are oblivious. We randomly select 5 users and use stall event as the KQI metric to evaluate the users' perceived experience. Moreover, the Pearson correlation coefficient between download ratio and downlink throughput is adopted to reveal the diversity of user engagement for different users. As shown in TABLE~\ref{table2}, we can see that the KPI/KQI-driven CoMP outperforms the KPI-driven CoMP in the stall event. Additionally, the correlation coefficient of the KPI/KQI-driven CoMP is bigger and smoother than the KPI-driven CoMP, which indicates the KPI/KQI-driven CoMP can better utilize the network throughput.

\section{Conclusion}

In this article, we first introduced the background of CoMP system, and investigated how backhaul, CSI, and synchronization affect KPIs in CoMP. Then, we illustrated the relationship between KQIs and KPIs, and analyzed why conventional CoMP fails to guarantee the KQIs demands for each user. Finally, we proposed a KPI/KQI-driven framework for CoMP that can provide KQIs guarantee for each user in 5G. Instead of merely focusing the PHY/MAC system performance, KPI/KQI-driven CoMP exploits interdependencies between CoMP and user's perceived experience. We believe that the KPI/KQI-driven CoMP in 5G will be an appealing topic, and some open problems remain to be resolved.

\begin{itemize}

\item \textbf{KPI/KQI-driven clustring}. By performing the smaller size cooperation, CoMP clustering can effectively reduce the cost of backhaul, clock, and pilot\cite{filedCoMP}. Existing CoMP clustering approaches focus on maximizing KPIs with the limit of cost of cooperation, and they fail to provide the KQI guarantee for each user's perceived experience. How to jointly consider KPI requirements and KQI requirements in CoMP clustering schemes is still an open problem.

\item \textbf{KPI/KQI modeling}. It is difficult to establish a precise mathematical model between KPIs and KQIs. On the one hand, in CoMP, cooperations among cells lead to the interdependence of the KQIs and KPIs of different cells. On the other hand, 5G systems will support various applications, and thus need massive KQIs to evaluate users' perceived experience to different applications. Existing models are built on maximizing the average of all users' perceived experience, but they fail to characterize the difference of individual user. Thus, it is worth considering to utilize machine learning to reveal the complicated relationships between the KQIs and the KPIs for CoMP in 5G.

\end{itemize}


\end{document}